\documentclass[12pt,a4paper]{article}
\usepackage{graphicx}
\usepackage{rotating}
\usepackage{amsmath,amssymb}
\begin{document}
\title{Cascade Model of an Anomaly \\ in Blazar Spectra at Very High Energy}
\author{Timur Dzhatdoev \\
Skobeltsyn Institute of Nuclear Physics, Lomonosov Moscow State University, \\ Leninskie gory 1-2, 119991 Moscow, Russia \\
timur1606@gmail.com}
\maketitle

\begin{abstract}
It is well known that the effect of gamma-ray absorption on extragalactic background light (EBL) is weakly expressed in the spectra of some blazars. It is shown that a secondary component generated by electromagnetic cascades might considerably decrease the statistical significance of this anomaly. Observational results indicate the existence of the cascade component in the spectra of extragalactic gamma-ray sources, thus supporting the proposed model.
\end{abstract}


\section{Introduction \label{sec:intr}}

For a long time it was expected that gamma-ray spectrum of a distant source in the very high energy (VHE, $E>$100 $GeV$) range would have a steepening resulting from the $\gamma\gamma \rightarrow e^{+}e^{-}$ absorption \cite{nik62}--\cite{gou67} on the extragalactic background light (EBL) photons at the energy region where the optical depth of this process $\tau_{\gamma\gamma}=1$. However, for some blazars --- gamma-ray loud active galactic nuclei --- the ratio of observed to expected intensity $K$ after accounting for the EBL absorption effect rises with $\tau_{\gamma\gamma}$; $K$ is larger for $\tau_{\gamma\gamma}>2$ than for $1<\tau_{\gamma\gamma}<2$ \cite{hor12}. The statistical significance of this anomaly is~4.2~$\sigma$~\cite{hor12}.

An effect of such kind is under discussion for about 15 years \cite{aha99}--\cite{pro00}; during this period of time two popular models of the anomaly were developed. The first model assumes that blazars are sources of hadrons that produce photons on the way from the source to the observer \cite{ury98}--\cite{ess10}. In this case a good collimation of the hadronic beam is required (the same as for the gamma-ray beam, i.e. $\theta_{Jet} \sim 1^{\circ}$); this condition is hard to meet, as after the exit from the blazar's jet a region of chaotic magnetic field is likely to exist, and this magnetic field is able to scatter the accelerated hadrons. The second model (e.g., \cite{san09}) postulates oscillations of photons to light exotic particles in magnetic field (and vice versa); the efficiency of the reverse conversion to photons is required to be quite high, independently of the direction to the source, and, therefore, of the Galactic magnetic field parameters in this direction. 

Another possible mechanism that could explain deviation of $K$ in the optically thick regime from the expected value is production of secondary particles in electromagnetic cascades \cite{aha99}, \cite{aha02}--\cite{ave07}. The present work is devoted to examination of the cascade component' influence to the statistical significance of the discussed anomaly.

\section{Observational data and analysis methods \label{sec:obsd}}

The present work is based on the sample of objects that were observed by Cherenkov telescopes listed in \cite{hor12}. The following spectra were selected, that contain: 1) at least 6 bins 2) at least one bin in the $\tau_{\gamma\gamma}>2$ region, in total 6 objects:  Mkn 421 \cite{tlu11}, Mkn 501 \cite{aha99}, H 1426+428 \cite{aha03}, 1ES 1101-232 \cite{aha06}, 1ES 0347-121 \cite{aha07}, 1ES 0414+009 \cite{abr12} (redshifts $z$ from $0.031$ to $0.287$).

The full MC calculation of $\sim10^{6}$ cascade spectra from primary photons in the energy range $E_{0}$= 100 GeV--100 TeV was performed with the ELMAG 2.02 code \cite{kac12}, assuming the EBL model of \cite{kne10}. The shape of the primary spectrum was set to \newline $J= C\cdot E_{0}^{-\gamma}\cdot exp(-E_{0}/E_{c})$, where $C$ is some constant, $\gamma$ --- the primary spectrum index, $E_{c}$ --- the energy corresponding to the end region of the spectrum. The 1D assumption used in the calculations is well justified, as in the energy range under investigation the angular radius of a typical cascade $\theta_{C}<< \theta_{Jet}$.

Various hypotheses were tested for the described sample of objects; for each hypothesis the corresponding $p$-value was calculated and converted to the equivalent statistical significance ($Z$-value, see \cite{cow11}, Chapter 2). For the significance calculation the values of the observed and model intensity in certain ranges of the $\tau_{\gamma\gamma}$ value were utilized (\cite{cow11}, Chapter 5.1). It was assumed that the distributions of the measured values in both directions (increasing and decreasing one) are Gaussian, but with the different parameters; a more comprehensive analysis calls for some more general, asymmetric distribution (e.g. \cite{kir03a}--\cite{kir03b}).

\section{The anomaly without the cascade component \label{sec:anom}}

\begin{figure}[t]
\centerline{\includegraphics[width=5.0in]{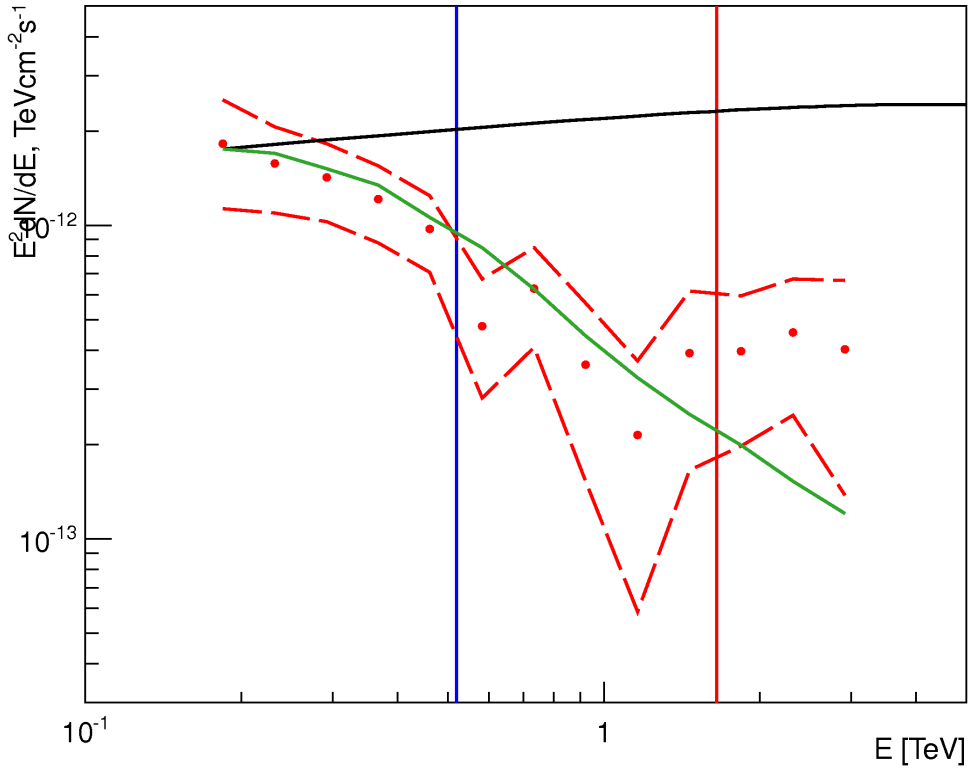}}
\caption{Illustration of a spectral energy distribution analysis for 1ES 1101-232 (z= 0.186) \cite{aha06} without the cascade component. Red circles --- the measured spectrum, red dashed line --- the total uncertainty of the measured spectrum $\sqrt{\sigma_{stat.}^{2}+\sigma_{syst.}^{2}}$ with account of statistical $\sigma_{stat.}$ and systematic $\sigma_{syst.}$ uncertainties, black line --- the primary model spectrum, green line --- the model of the observed spectrum, blue vertical line --- the energy, at which $\tau_{\gamma\gamma}=1$, blue vertical line --- the same, but for $\tau_{\gamma\gamma}=2$.}
\label{Fig01}
\end{figure}

If cascade electrons are deflected by the extragalactic magnetic field (EGMF) to the angles $\theta_{M}>>\theta_{Jet}$, the cascade component in the observed spectrum is practically absent. An example of the spectral energy distribution (SED, $E^{2}dN/dE$) for one of the analysed spectra in this case is shown in Fig.~\ref{Fig01}. The model of the observed spectrum (green line) doesn't contradict the observations at the $\tau_{\gamma\gamma}<2$ region, but for $\tau_{\gamma\gamma}>2$ the model intensity is somewhat lower than needed to desribe the data.

The hypothesis that without the cascade component the model is still able to describe the data in the $\tau_{\gamma\gamma}>2$ region for the sample of 6 objects was rejected at the statistical significance level $Z_{A}=3.1 \sigma$. A similar result was obtained in \cite{hor12} with a different method: $Z= 3.8 \sigma$. In general, these results are in good coincidence, because the systematic uncertainty of the spectral shape that was accounted for in the present work tends to decrease the significance.

\section{Influence of the cascade component to the statistical significance of the anomaly \label{sec:casc}}

\begin{figure}[t]
\centerline{\includegraphics[width=6.0in]{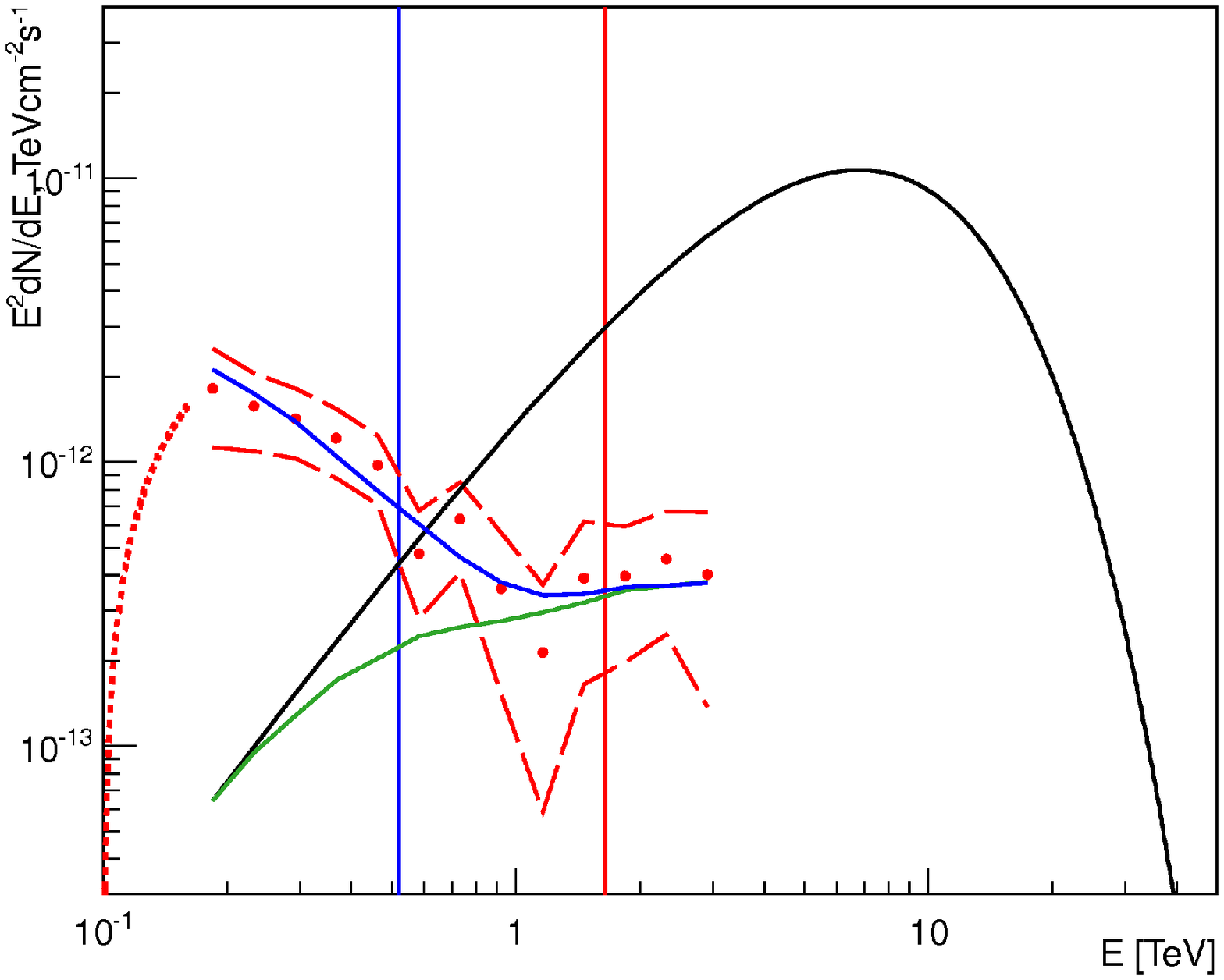}}
\caption{Analysis of the same spectrum as in Fig.~\ref{Fig01}, but with account of the cascade component. The notations are the same as in Fig.~\ref{Fig01}, except: 1. blue curve that denotes the model spectrum with account of the cascade component 2. red dashed curve that qualitatively illustrates a possible influence of EGMF on the cascade component spectrum.}
\label{Fig02}
\end{figure}

For the EGMF strength $B<$10$^{-16}$ $G$ and correlation length $L_{c}$= 1 $Mpc$, cascade photons can significantly contribute to observed spectrum at $E>$ 100 $GeV$ (see Fig.~\ref{Fig02}). While the $\gamma\gamma \rightarrow e^{+}e^{-}$ process due to its threshold nature occurs only at EBL photons for $E_{\gamma}<$100 $TeV$, the main target for electrons in the Inverse Compton (IC) interaction acts is dense cosmic microwave background (CMB) \cite{aha02}, \cite{kac12}, whose photons have comparatively low energy. Therefore, the IC process, as a rule, occurs in the Thomson regime, thus the photons of the next generation carry only a small fraction of the primary electron energy. This peculiarity of the cascade process is clearly seen in Fig.~\ref{Fig02}: the secondary component from cascade photons (blue curve) gives the main contribution at comparatively low energy, in the optically thin regime. 

In the framework of the presented model, the contribution of the cascade component at $\tau_{\gamma\gamma}>1$ cannot account for the anomaly in the distant ($z>$0.2) blazar spectra at $E>$ 1 $TeV$. However, the secondary photons do influence the interpretation of the data indirectly: for better agreement of the model with observations, the primary spectrum in Fig.~\ref{Fig02} is chosen harder, than in Fig.~\ref{Fig01}; this circumstance, in its turn, leads to a significant increase of the model intensity at the $\tau_{\gamma\gamma}>2$ region. An analysis similar to the one conducted in sec.~\ref{sec:anom}, but with account of the cascade component, yields $Z_{A}$= 0.36 $\sigma$ (the anomaly is practically absent).

\section{Discussion \label{sec:disc}}

The analysis of the cascade component influence on the gamma-ray spectrum in the present work was carried out assuming $B<<$10$^{-16}$ $G$. The probable influence of the EGMF on the cascade component spectrum is schematically shown by dashed red curve in Fig.~\ref{Fig02}. A similar spectral shape was already observed during a flare of Mkn 501 \cite{ner12}, that was simultaneously observed by the orbital Fermi LAT instrument \cite{atw09} and the VERITAS Cherenkov telescope \cite{abd11}. Additionally, in \cite{ner12} a constraint on the impulse front rise time $T_F<$10 $days$ (90 \% C.L.) at E$\sim$100 GeV was obtained, as well as an estimate of $B$= 10$^{-16}$--10$^{-17}$ G for $L_c$= 1 Mpc.

If $B$ and $L_c$ are nearly the same for all 6 objects of the considered sample, the characteristic variability time of the cascade component for them could be estimated as \cite{aha99} $(z/z(Mkn501))\cdot T_{F}$ = 10-80 $days$, depending on $z$. For the two nearest objects of the sample the contribution of the cascade component is small; 4 blazars with $z>0.1$ do not show variability with period less than several months. This doesn't allow to set additional constraints on the cascade component' contribution using temporal information.

Existing bounds on $B$ (e.g., \cite{abr14}, where the angular distribution of observed photons, that depends on the cascade component contribution, is analysed) doesn't contradict the assumptions of the considered model. Future instruments with either high sensitivity (e.g. CTA \cite{ach13}) or extremely good angular resolution (e.g. emulsion gamma-ray telescope \cite{aok12} or the GAMMA-400 apparatus \cite{gal13}) would possibly allow to measure the EGMF strength. Finally, the recent indication that the directions to the objects observed by Cherenkov telescopes and the Fermi LAT instrument are predominantly point to large-scale voids \cite{fur14}, and, therefore, to low $B$ regions, is a direct argument in favor of existence of the cascade component in blazar spectra.

\section{Conclusions \label{sec:conc}}

In the present work, for the first time, a possibility to explain the anomaly in blazar spectra at VHE region in the framework of cascade model is considered in detail (i.e. quantitatively). For the first time, it is shown that the influence of secondary photons that might decrease the statistical significance of the anomaly may have an indirect nature: while these photons contribute mainly in the optically thin regime, the cascade component allows the harder primary spectrum to fit the data, which leads to better agreement of the model with observations in the optically thick regime.

The cascade model doesn't require a single additional assumption and doesn't contradict to observations of Cherenkov and orbital gamma-ray telescopes; at the present time this model is the simplest explanation of the anomaly for some blazars.

\textit{Additional note. An evidence for broadened angular distribution at $\sim$1 GeV for a sample of low redshift blazars was recently found in \cite{che14} using Fermi LAT data and interpreted as a proof of pair halos existence; an estimate of EGMF strength of 10$^{-15}$--10$^{-17}$ G was obtained. The latter value doesn't contradict the model presented here.}

\section{Acknowledgements \label{sec:ackw}}

The calculations were performed using the SINP MSU OOKM (the Center for Operative Cosmic Data Monitoring) computer cluster; the author is grateful to V.V. Kalegaev for permission to use the hardware, and to V.O. Barinova, M.D. Nguen, D.A. Parunakyan for technical support. The work was partially supported by the Russian President grant NS 3110.2014.2.

\end{document}